\documentstyle[aps,prl,epsfig,floats]{revtex}
\begin{document}
\draft
\twocolumn[\hsize\textwidth\columnwidth\hsize\csname @twocolumnfalse\endcsname

\title{Fluctuation-Facilitated Charge Migration along DNA}

\author{R.Bruinsma, G.Gruner, M.R.D'Orsogna and J.Rudnick}

\address{Physics Department, University of California, Los Angeles, CA
	90095-1547}
\maketitle
                
\begin{abstract}
We propose a model Hamiltonian for charge transfer along the DNA double helix
with temperature driven fluctuations in the base pair positions acting as 
the rate limiting factor for charge transfer between neighboring base pairs. 
We compare the predictions of the model with the recent work of J.K. Barton
and A.H. Zewail (Proc.~Natl.~Acad.~Sci.~USA, {\bf 96},
6014 (1999)) on the unusual two-stage charge transfer of DNA.

\par 
\end{abstract}
]

\makeatletter
\global\@specialpagefalse
\let\@evenhead\@oddhead
\makeatother     
    
\par

Charge transport and electrical conduction are known to occur in a wide range
of organic linear chain crystals of stacked planar molecules 
\cite{jerome}.
Transfer rates from molecule to molecule are determined by the 
single-particle integral $\tau$, with typical rates of the order of 
10$^{15}$ sec$^{-1}$.
Strong interaction between the electronic degrees of freedom and molecular 
vibrations may reduce this to 10$^{12}$ sec$^{-1}$, a typical lattice or 
intramolecular vibration frequency. 
By comparison, biochemical charge transfer
processes, such as those encountered in the metabolic redox 
(oxydation-reduction) chains \cite{stryer}, usually are much slower
(down to 1 sec). Key steps often involve some form of large scale motion
of the molecule. 

\par
DNA can be considered as a one-dimensional, a-periodic, linear chain of 
stacked base-pairs. 
More than 30 years ago it was suggested \cite{eley}
that duplex DNA might support electron transport in a manner similar to 
that of linear chain compounds, namely by tunneling along overlapping
$\pi$ orbitals located on the base-pairs. 
Barton et al.\cite{barton93}
first presented evidence that photo-induced, radical cations can travel along 
DNA molecules in aqueous solution over quite considerable distances (more 
than 40 \AA, or about ten base-pairs). 
If so, DNA might present us
with  flexible, molecular-size wires able to transport charge in aqueous 
environments. 
Possible applications range from micro-electronics to 
long range detection of DNA damage. 
Subsequent studies by a number of
groups reported a wide range of values for the effective inverse spatial
carrier decay length $\beta$, ranging from as low \cite{henderson} as 
0.02 \AA $^{-1}$ to as high as around one \AA$ ^{-1}$, 
the different values most probably reflecting differences in charge transfer
for different base sequences \cite{giese}.

\par
Recently, Barton and Zewail \cite{barton99} (BZ) used femtosecond spectroscopy
to measure the rates of the DNA charge transfer process. 
An unusual two-step
decay process was observed with characteristic time scales of 5 and 75
ps respectively. 
Ab-initio molecular-orbital calculations
\cite{saito1} find that DNA has a large single particle band-gap,
and a transfer integral $\tau$ of order $0.1$ eV. 
This would lead to a charge 
transfer rate $\tau / h$ for coherent tunneling that is comparable
to that of the linear chain compounds, but that is much too high compared to
the rates measured by BZ.

\par

Apart from coherent tunneling \cite{eley}, a number of alternative transport
mechanisms have been proposed, in particular incoherent, phonon-assisted 
electron hopping between bases, with the electron wave fully localized
on each subsequent base-pair \cite{ly,jortner}. 
This would reduce 
the transfer rate to a typical intra-molecular vibrational frequency 
(ps$^{-1}$), but this is still much too large to explain the slow second 
stage step of the decay. 
It was also suggested \cite{henderson} that 
a charged 
radical could induce a polaronic distortion \cite{emin} of the DNA internal
structure that might control charge transfer. 
The explanation proposed by BZ
for the long relaxation times is that large amplitude thermal fluctuations
of the intercalated photoreceptor sets the rate limiting step for the
charge transfer.

\par
The aim of this letter is to construct a model Hamiltonian to treat charge 
transfer along a chain under conditions of large structural fluctuations,
and suggest that thermally induced structural disorder interferes with 
the $\pi$ orbital overlap mediated charge transfer, leading to long 
relaxation times. 
To construct this Hamiltonian, we first must discuss the origin of the 
structural fluctuations. Figure 1 shows an example of a typical DNA 
configuration obtained by a molecular dynamics (MD) simulation 
\cite{swaminathan}. The relative orientation of neighboring bases along DNA
is characterized by a set of collective variables such as the relative
roll and twist angles (R and T), and the relative slide displacement
(see Fig.1). 
Long time MD simulations of DNA lead to typical RMS fluctuation angles
for R and T of order 5 and 9 degrees \cite{rosenberg} in the ps to ns 
time-window, while the mean base-pair spacing also shows large amplitude 
fluctuations \cite{note1}. 
Structural fluctuations in the ps to ns 
time-window have been observed \cite{brauns} experimentally as dynamic
Stokes shifts in the fluorescence spectrum of the DNA. 
The local fluctuations
are extraordinarily strong compared with those due to thermally excited 
phonon modes in crystalline linear chain materials. 
The unusual `softness' 
of the R,S, and T variables is also reflected by the fact that their mean 
values vary greatly depending on base-pair sequence \cite{calladine:book}.

\par
In our simplified model, we include only two collective modes. 
The first mode is an angular variable $\theta(t)$, which is that relative 
rotation angle of the two bases which couples most efficiently to the
$\pi$-orbital tunnel matrix element.
Next, the displacement variable $y(t)$ represents that collective mode      
which couples most efficiently to the
%___________________________
%_________________________________________________________^M
\begin{figure}
 \vbox to 9.5cm {\vss\hbox to 9cm
 {\hss\
   {\includegraphics{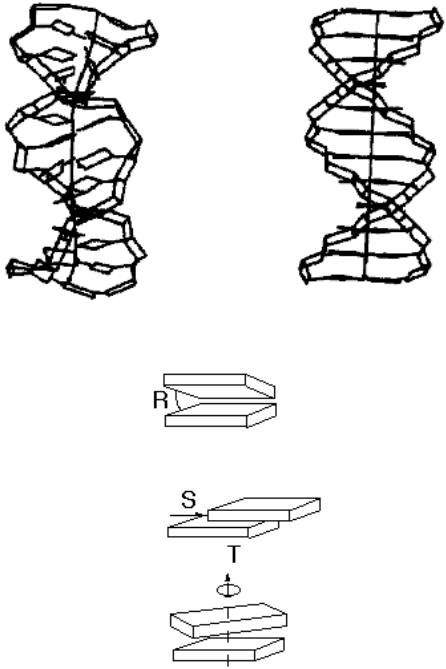}
   }
  \hss}
 }
\vspace{0.4cm}
\caption{ {\it Top:} Snapshot of the result of a 140 ps room temperature MD simulation
of DNA, by Swaminathan  (left panel) as
compared to the ordered DNA structure at $t=0$ (right panel).
{\it Bottom:} Schematization of the roll and twist angles R and T and of 
the relative
slide displacement variable S. 
}
\end{figure}
%____________________________________________________________      
\noindent
%________________________________________
on-site energy of the radical.
This second form of coupling provides the necessary mechanism for 
energy transfer between the charge and the the thermal reservoir required 
for hopping transport along a random sequence of base-pairs with different 
on-site energies. 
Both $\theta(t)$ and $y(t)$ are treated as classical 
harmonic variables that are coupled to a heat bath of oscillators.
As a result, $\theta(t)$ and $y(t)$ obey -in the absence of the radical-
the following Langevin equations:

\begin{eqnarray}
\nonumber
\displaystyle
M\ddot{y} + \gamma_y \dot{y} + M\Omega^2_y y = \eta_y (t),      \\
I\ddot{\theta} + \gamma_\theta \dot{\theta} + I \Omega_\theta^2 \theta  
			= \eta_\theta (t). \label{langevin} 
\end{eqnarray}
\noindent
In Eq.~\ref{langevin}, $I$ is the reduced moment of inertia for 
the relative rotation 
of the two adjacent bases, $\Omega_{\theta}$ is the oscillator frequency 
of the rotation mode, while $M$ and $\Omega_{y}$ are the reduced mass and
natural frequency of the displacement mode. 
The values of $M$,$I$,
$\Omega_{\theta}$,$\Omega_{y}$, and the damping coefficients are obtained
by comparing the Fourier power spectra of $y(t)$ and $\theta(t)$ obtained 
from Eq.~\ref{langevin}, 
to power spectra of MD simulations of DNA \cite{note2}.
From a typical long time (10 ns) MD series, we find oscillation periods of 
$2\pi / \Omega$ of order 1-10 ps, a (large) mass $M$ of order 1-10 kDalton,
(1~Dalton~=~1~a.m.u.) a moment of inertia $I$ of order 
$10^{2} ~ k_{B} \Omega_{\theta}^{-2}$,
and relaxation times comparable to the oscillation period (i.e. the slow
modes are close to critical damping). 
The amplitudes of the white-noise
variables $\eta(t)$ follow from the fluctuation-dissipation theorem
for classical variables. 
We will neglect mode coupling between different 
pairs of adjacent base-pairs. 
The two modes are then coupled to a 
one-dimensional, tight-binding Hamiltonian for single particle charge 
transport:

\begin{eqnarray}
\nonumber
H = \sum_i \{ \tau(\theta_{i,i+1}) (c^{\dagger}_{i+1}c_i + c^{\dagger}_i
	c_{i+1})
    + \epsilon_i c^{\dagger}_i c_i  \\ \nonumber 
+{1 \over 2} I_i (\dot{\theta}^2_{i,i+1} + \Omega^2_{\theta,i} \theta^2_{i,i+1})\\
+ { 1 \over 2} M_i (\dot{y}_i^2 + \Omega^2_{y,i} (y_i + y_{0,i} c^\dagger_i c_i)^2)\}
+ H_{bath}(\{\theta,y\}).
\label{hamiltonian}
\end{eqnarray}

\noindent
In Eq.~\ref{hamiltonian}, 
$\epsilon_{i}$ is the on-site electronic energy. 
The distance
$y_{0,i}$ is the change in the equilibrium value of the $y$ variable of the 
i-th base when the particle localizes on that site, while 
$M\Omega^{2}y_{0}^{2}$ is the typical deformation energy. 

\par
Certain limiting cases of this general Hamiltonian are familiar from studies
of one-dimensional charge transport. 
For uniform $\epsilon_{i}$ and for fixed
$\theta$, $H$ is the Hamiltonian of a tight-binding polaron \cite{emin}.
For fixed $\theta$ and and $y$ and random $\epsilon_{i}$, $H$ is the Anderson
Hamiltonian for localization in one-dimension. 
For the case of DNA,
we assume that site-to-site differences in the value of $\epsilon_{i}$
are of order $0.1$eV based on the sequence dependent differences in the
ionization potential \cite{saito2}. 
Next, the transfer integral $\tau(\theta)$ will be assumed to be small 
compared to the thermal energy $k_{B}T$ for $\theta$ near a special 
value, denoted by $\theta^{\ast}$, the `rapid decay state'.
Finally, the characteristic interaction energy $M \Omega^{2}y_{0}^2$ between
the charged radical and the on-site structural variable $y$ is assumed to be
large compared to $k_{B}T$ (e.g. due to electrostatic effects) and of order
$\epsilon$.

\par
In this unusual, high temperature, strong coupling regime, the transfer 
integral $\tau(\theta)$ is the lowest energy scale. 
Under these conditions,
particle motion described by $H$ is indeed dominated by incoherent hopping 
from site to site.
We first restrict ourselves to the case of a particle which resides at site 
$A$ at time $t=0$ and then hops to the neighboring site $B$ with a different
on-site energy. 
For fixed $\theta$, the transition rate $\Gamma(\theta)$
for incoherent charge transfer between $A$ and $B$ can be computed by 
applying the method of Garg, Onuchic and Ambegaokar (GOA) \cite{garg}.
In Fig.2 we show the two potential energy surfaces $V^{+}(y)= 
\langle A|H|A \rangle$ and $V^{-}(y)=\langle B|H|B \rangle $ for the 
particle respectively on the $A$ and  $B$ sites as
a function of $y$. 
Efficient transfer between the two potential 
energy surfaces takes place nearly exclusively at the crossing points
$y^{\ast}$ where $V^{+}(y^{\ast})=V^{-}(y^{\ast})$, 
shown in Fig. 2. 
Note that
an energy barrier $E_{r}$ must be overcome to reach this crossing point.
In the high temperature, strong coupling limit, the on-site probability
decays exponentially with a rate:
\begin{equation}
\displaystyle
\Gamma(\theta) \simeq { \tau(\theta)^2 \over \hbar}
		\left[ {\pi \over E_r k_B T} \right]^{1/2} 
	exp (-{E_f / 
		k_B T}),
\label{probability}
\end{equation}
where the energy scale $E_{r}$ defined in Fig.2, depends on 
%____________________________________
%_______________________________________________________________________
\begin{figure}[tb]
% \vbox to 8 cm {\vss\hbox to 7cm
% {\hss\
%   {\special{psfile=fig2.eps angle=0
%               hscale=90 vscale=90
%                hoffset=-90 voffset=15
%                }
%   }
%  \hss}
% }
\centerline{\epsfig{file=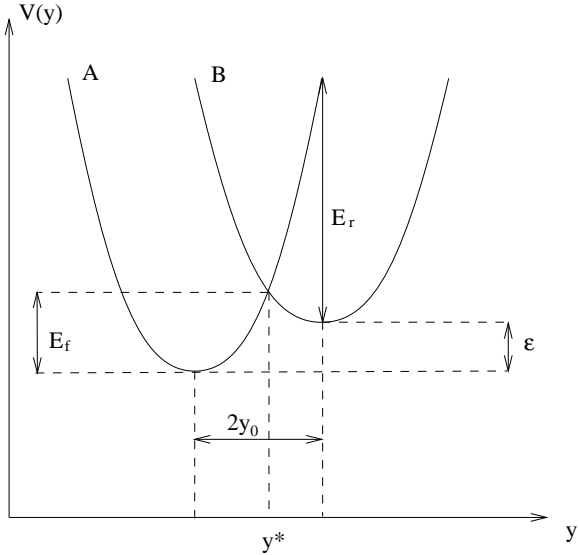, width=3in}}
\caption{Potential energy surfaces for a particle localized on either of two
neighboring sites A and B as a function of the collective variable $y$.
The $\theta$ variable is assumed fixed. Non-adiabatic transfer between the
two surfaces takes place exclusively at the degeneracy point $y^{\ast}$.
}
\end{figure}
%_____________________________________________________________________________
\noindent
%________________________________
the difference
between the on-site energies. 
The validity condition for 
Eq.~\ref{probability} is that 
$\tau(\theta)$ must be small compared 
to $(E_{r} k_{B}T)^{1/2}$.
By itself, Eq.~\ref{probability} 
does not account for a two stage decay, as observed by BZ.
In a typical ensemble of radical sites (with the same on-site energies)
there must be significant heterogeneity, with less likely states
characterized by low charge transfer rates. 
This heterogeneity
is incorporated by demanding that $\tau(\theta)$ is appreciable 
only for $\theta$ near the special value $\theta^{\ast}$, 
while tunneling plays 
no role for different $\theta$ values \cite{note3}. 
If $\theta$
undergoes large amplitude thermal fluctuations, then there should be 
considerable heterogeneity for the transfer rates.

\par

Assume then that, at time $t=0$, an ensemble of particles is prepared 
on the $A$ site, with the $\theta(0)$ variable obeying the Boltzmann 
distribution. 
We define the probability density $P(\theta, t) d\theta$ 
to be the fraction of radicals at time $t$ that are still on site $A$,
and whose $\theta$ values is in the range between $\theta$ and 
$\theta + d\theta$. 
For the overdamped case of Eq.~\ref{langevin}, 
$P(\theta, t)$
obeys the following equation:

\begin{eqnarray}
\nonumber
{\partial P(\theta,t) \over \partial t} =
{k_BT \over \gamma_\theta} { \partial^2 P(\theta,t) \over \partial\theta^2} 
+{1 \over \gamma_\theta} {\partial \over \partial \theta} \large[
 I\Omega_\theta^2 \theta P(\theta,t) \large] \\
- \Gamma(\theta) P(\theta,t).
\label{fokker} 
\end{eqnarray}
\noindent
We can now use Eq.~\ref{fokker} 
to discuss the decay rate. 
For $\Gamma(\theta^{\ast})$
large compared to the thermal equilibration rate $\tau_{\theta}^{-1}$
two forms of decay are encountered \cite{alexander}.

\par
{\it (i) Early stage decay:} 
At times $t=0$, a certain fraction of the oscillators has an energy 
exceeding $E=(1/2) I (\Omega_{\theta} \theta^{\ast})^2$. 
These oscillators will 
pass through the $\theta=\theta^{\ast}$ point within a time of order
max$ \{ \tau_{\theta},\Omega_{\theta}^{-1} \}$.
When this happens, there is a finite probability for charge transfer to 
take place. On a time scale of order 
max$ \{ \tau_{\theta},\Omega_{\theta}^{-1} \}$, these high energy oscillators
are removed from the probability distribuition.

\par
\par
{\it (ii) Late stage decay:}
After the high energy oscillators have been removed from the distribution,
further decay requires energy diffusion along the oscillator scale
from lower energies towards 
$E(\theta^{\ast})=(1/2) I (\Omega_{\theta} \theta^{\ast})^2$. 
Once the energy of 
an oscillator reaches this value, efficient charge transfer takes place.
After a standard, but lenghty, analysis of Eq.~\ref{fokker} 
we find that the late 
stage decay rate is:

\begin{equation}
\displaystyle
k_{LATE} = \tau_\theta \Omega_\theta^2 \left[
{\theta^\ast e^{- {1 \over 2} \beta I \Omega_\theta^2 {\theta^\ast}^2}
\over
\int_{-\infty}^\infty 
 e^{- {1 \over 2} \beta I \Omega_\theta^2 {\theta^\ast}^2} d\theta} \right].
\label{klate}
\end{equation}
\noindent
The factor in front of Eq.~\ref{klate} is of the same order as the early 
stage decay rate
while the term in brackets is of the order of the thermal probability that 
$\theta$ exceeds $\theta^{\ast}$. 
The second stage decay rate strongly increases
with increasing temperature, while the early stage decay rate is not expected
to be strongly temperature dependent although the fraction of sites that
exhibits early stage decay should be strongly temperature dependent
(of the same order of the term in brackets of Eq.~\ref{klate}).

\par
The time dependence of $P(\theta,t)$ would be consistent with 
the observations of BZ if this thermal probability of the $\theta^\ast$ state
is of order $10^{-2}$. 
In that case, one percent of the sites would show rapid
decay with time scales of order ps, while the remainder would show decay
slowed down by a factor of $10^{2}$. 
We only treated here the nearest 
neighbor hopping process. 
Charge transport over longer distances described by
our Hamiltonian reduces - under the assumed conditions - to a classical 
one-dimensional diffusion in a random medium with site specific transfer 
rates. 
The transport properties of such systems have been extensively 
discussed elsewhere \cite{alexander}.

\par 
In conclusion, we propose that charge transport along DNA proceeds by 
classical diffusion with high amplitude thermal fluctuations providing 
the rate limiting step for the site-to-site charge transfer. If correct,
charge transport along DNA would have unique characteristics as compared 
to linear chain compounds. 
Since the radical severly deforms the local 
structure it might be considered as a polaron in the strong coupling limit
but, unlike polaronic transport, hopping is controlled by thermal 
fluctuations. Indeed, Eq.~\ref{klate} 
predicts that the charge transfer rate should 
strongly increase with temperature which is consitent with the observations of
BZ. 
A better description of the mechanism proposed in this paper for 
charge transfer along DNA would be to consider it as a repeated sequence of
reversible oxydation-reduction reactions. 
The site-to-site 
charge transfer would be viewed as a chemical reaction dominated by a 
transition state where the collective variables $y$ and $\theta$
assume a special value ($y^{\ast}$ and  $\theta^{\ast}$ respectively).
We are not aware of any of the linear chain compounds exhibiting this 
curious form of charge transfer. 
On the other hand, a recent single-molecule 
optical study of a particular reversible oxydation-reduction reaction did 
report \cite{lu} two-stage non-exponential behavior but with decay rates 
much lower than those measured by BZ (in the range of 1 sec$^{-1}$).

The higher rates of charge transfer in DNA would be due to the fact that 
the molecular motion of the bases still is significantly restrained by the 
backbone. 
If the present analysis is appropriate, then charge transport in
DNA occupies a unique position intermediate between charge transport 
in solid-state materials and charge transport in biochemical charge transfer
reactions.

\par

Finally, the proposed Hamiltonian obviously incorporates a number of rather
serious simplifications. 
We include the collective modes only in a 
schematic way. 
We require a large structural on-site distortion
of a site by the particle (of order $0.1$ eV), but this is likely to
require an anharmonic description of the collective modes.
We did not include coupling of modes of adjacent pairs, the double-stranded
nature of DNA with the possibility of inter-chain
charge transfer or effects related to the tertiary structure, i.e. the coiling
of the duplex.

\par
We would like to thank J.Barton and F.Pincus for stimulating and useful 
discussions. We would like to thank J.Onuchic for providing a copy of his 
thesis and for discussions. RB would like to thank the Rothschild 
Foundation and NSF Grant DMR-9407741 for financial support.

%____________________________________
 
\par

\narrowtext

%_______________________________________________________________________

%________________________________________________________________________

\begin{references}
\bibitem{note1}
	Structural transitions, from the B to the A form of DNA, 
	are encountered on nanosecond timescales in MD simulations.
\bibitem{note2}
	Using the fact that both $I \Omega^{2}_{\theta} \langle \theta^{2}
	\rangle$, and $I \langle (d\theta/dt)^2 \rangle$ must equal
 	$k_{B}T$.
\bibitem{note3}
	The transfer integral was computed in \cite{saito2} as a function of
	the relative twist $T$ and the calculated function $\tau (T)$
	and it exhibits  multiple maxima.
\bibitem{note4}
	If $\Gamma(\Theta^{\ast})$ is small compared to the thermal 
	equilibration rate, then the probability distribution retains the
	Boltzmann form with a time dependent prefactor that decays as a
	single exponential with a rate equal to the Boltzmann weighted
	average of $\Gamma(\theta)$.
\end{references}
\end{document}